\documentclass[twocolumn,showpacs,preprintnumbers,amsmath,amssymb,prl]{revtex4}

\usepackage{graphicx}
\usepackage{dcolumn}
\usepackage{bm}

\begin{document}

\preprint{}

\title{Bipolariton laser emission from a GaAs microcavity}

\author{L. M. Moreira, J. C. Gonzalez, A. G. Oliveira and F. M.
Matinaga}\thanks{matinaga@fisica.ufmg.br} \affiliation{Departamento
de F\'{\i}sica, Universidade Federal de Minas Gerais, Belo
Horizonte, MG, 30123-970 Brazil\\}



\begin{abstract}

Biexciton emission properties were studied in a single GaAs quantum
well semiconductor planar microcavity by photoluminescence
measurements at low temperatures. At high pump intensity a
bipolariton emission appears close to the lower polariton mode. This
new mode appears when we detune the cavity resonance out of the
lower polariton branch, showing a laser like behavior. Very small
lines widths were measured, lying below 110 $\mu$eV and 150 $\mu$eV
for polariton and bipolariton emission respectively. The
input/output power (I/O) measurements show that the bipolariton
emission has a weaker coupling efficiency compared to previous
results for polariton emission. Simultaneous photoluminescence and
near field measurements show that the polariton and bipolariton
emission are spectrally and spatially separated.

\end{abstract}

\maketitle

\section{I.Introduction}

In recent years the possibility to realize low-dimensional
heterostructures such as quantum wells (QW) enhanced the Coulomb
interactions between electrons and holes in these structures. The
interaction of excitons with photons is also increased with the
spatial confinement in low-dimensional systems. Recent development
of microcavity structures\cite{skolnick} enhanced the interaction of
photons and excitons producing exciton-polariton quasi-particles
(polariton)\cite{savona1}. The interaction between single excitons
is also possible, creating biexcitons or excitonic molecules
\cite{ivanov} and therefore, the bipolariton particle in a
microcavity \cite{baars}.

The semiconductor planar microcavity (SMC) is a Fabry-Perot cavity
formed by two distributed Bragg reflectors (DBR) separated by a
multiple integer of $\lambda$/2 distance.  Quantum wells are located
at the anti nodes position of the cavity electromagnetic field. The
photons are confined inside the cavity in the grown direction,
modifying the energy dispersion relationship to a parabolic form.
Tuning the photon energy inside the cavity is possible to resonantly
excite QW excitons and create polaritons states distributed in the
upper and lower polariton branches (UP and LP). In a high Q
microcavity, the coupling between cavity photons and excitons leads
to a strong coupling regime in the SMC identified by the Rabi
splitting with frequency $\Omega$ \cite{weisbuch}. One of the most
intriguing characteristics is the polaritons ´s integer spin,
obeying Bose-Einstein statistic \cite{moskalenko}, consequently the
polaritons could condensate to a final state \cite{deng}. An other
interesting effect in a SMC, polaritons created at high pump
intensity with a parallel wave vector ($k_{\|}$), such that the LP
population per mode reaches one, can be scattered elastically into
two polaritons, one in the bottom of the LP branch (signal)and the
other one to a higher energy side (idler). This mechanism preserves
phase coherence and a high intensity emission signal for the two
correlated beams, besides a laser like source light \cite{koch}.

Neukirch et al. addressed one of the first experimental evidence of
biexcitons in a SMC \cite{neukirch}. By using a II-VI SMC, they
could distinguish between bipolaritons and polaritons resonance in
pump-probe experiments. Several other groups followed studying the
properties of bipolaritons in SMC using mainly pump-probe
\cite{borri} and four-wave mixing \cite{ivanov2} experiments. Baars
et al. \cite{baars} mapped the energy dispersion relationship for
the bipolariton and compared it with calculations described by $4x4$
matrix Hamiltonian for a coupled four-level model. The correct
description of the biexcitons formation in the SMC comes with the
bipolariton formalism that leads to similar polariton dispersion
curves, showing an lower and upper bipolariton branch (LB and UB)
associated with a splitting frequency.

In this paper we report optical characteristics of a bipolariton
emission in a GaAs QW SMC. Our results show laser like linewidth
narrowing in the bipolariton emissions, and also a emission
intensity relation with excitation power (I/O data). The bipolariton
emission was observed to be blue or red shifted from the LP
depending of the detuning ($\Delta=E_{c} - E_{x}$, where $E_{c}$ is
the cavity resonance energy and $E_{x}$ the exciton energy). The
emission far field image shows two emissions pattern, spatially and
spectrally separated.

\section{II.Experimental Details}

The SMC used in this experiment was grown by molecular beam epitaxy
(MBE) and consists of a single $100 Å$ GaAs QW in the middle of a
$\lambda/n$ (where n stands for the refractive index of the layer)
cavity of $Al_{0.3}Ga_{0.7}As$, sandwiched between 24 (top) and 29.5
(bottom) pairs of distributed Brag reflectors
$Al_{0.2}Ga_{0.8}As/AlAs$. The microcavity was designed to operate
at $\lambda=800$ nm which matches the QW emission when cooled to 7
K. The cavity length varies across the sample position, which allows
positive and negative detuning with respect to the match between the
cavity resonance and the exciton emission. More details on the
sample can be found in Ref. \cite{cotta}.

The experiment was performed using a liquid helium cold finger
cryostat held at a temperature of 7 K. A continuous wave (CW)
Ti:Sapphire tunable laser with an variable angle of incidence with
respect to the cavity normal direction was used to excite the
sample. The emission was collected from the sample in a
backscattered geometry and dispersed by the $1800$ gr/mm grating
spectrometer and detected by a charge couple device (CCD) camera.
For the near field images, we used a second camera in front of the
emission that could be easily removed for the spectra measurements.

\section{III.Results and Discussion}

\begin{figure}
\includegraphics[angle = 0, scale = 4.4]{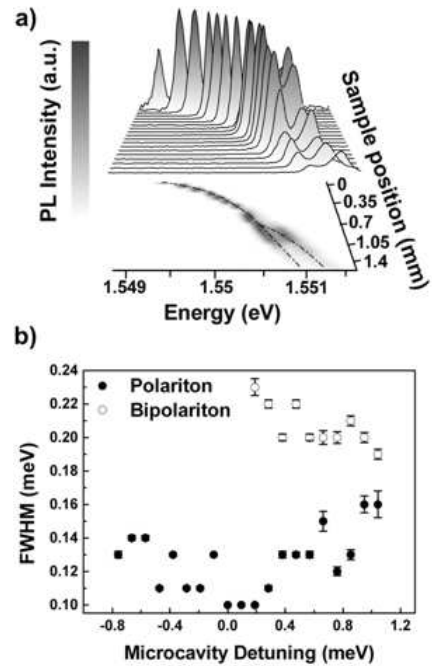}
\caption{(a) The PL spectra of the LP and LB as a function of sample
position and the projection of the intensity spectra. (b) The FWHM
for both emissions as a function of the detuning.}\label{mapa}
\end{figure}

The LP emission was observed pumping the SMC at $k_{\|}=7.7x10^{4}$
$cm^{-1}$ with an energy $\Omega$=3.2 meV above the lower polariton
branch. Out of this optimum condition, by detuning the cavity
resonance ($\Delta>0$), strong LB emission was induced close to the
LP peak with 400 mW pump power, as illustrated in Fig. \ref{mapa}.
We could see one asymmetry in the polariton laser spectrum for small
blue shift $\Delta$, which evolved to a two separate line emission
(~0.25 meV) as $\Delta$ increases. The figure shows a series of
photoluminescence spectrum normalized with respect to the
backscattered laser intensity for a fixed 400 mW pump power. The
total $\Delta$ shift corresponds to a range of 1.9 meV or a 1.4 mm
shift in cavity position. The LB mode has a higher energy in
relation to LP emission and the energy difference between them
increases while $\Delta$ is shifted for higher energies. Fig.
\ref{mapa}(b) shows the full width at half maximum (FWHM) of the
photoluminescence (Fig. \ref{mapa}(a)) as a function of the
detuning. This linewidth behavior shows the LB formation process
(decreasing) while the LP FWHM increases indicate the LP laser
efficiency loss due to the higher.

\begin{figure}
\includegraphics[angle = 0, scale = 5.5]{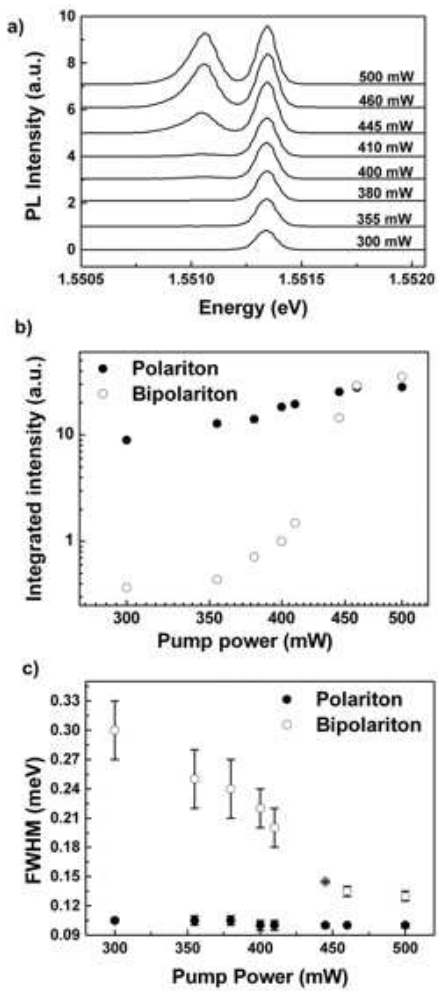}
\caption{(a) LP and LB PL spectra as a function of pump power.(b)
The FWHM behavior for the two emission peaks. (c) Integrated power
relation.} \label{potencia}
\end{figure}

In Fig. \ref{potencia} we observe the LB behavior by lowering the
pump power for a fixed detuning. The LB mode appear in this
measurement $~0.27$ meV red shifted from the LP mode (Fig.
\ref{potencia}(a)), contrary to the observed shift in the previous
experiment (Fig. \ref{mapa}).  The integrated intensity relation I/O
shows a transition from spontaneous LB emission to a laser like
behavior around $400$ mW pump power threshold (Pth). This transition
is also observed in Fig. \ref{mapa}(c) for the FWHM of both
emissions as function of the pump power. At $400$ mW the LB FWHM
decreases rapidly and become stationary at 500 mW with a value of
~130 $\mu$eV. The usual kink in the linewidth and the I/O curve at
Pth is not clear enough here because the LB mode is too close the LP
mode and we could not separate both spectra clearly for lower
excitation pump power. The LP emission that was always above the
threshold has a value of $~100$ $\mu$eV FWHM whose behavior have
been described by Cotta et al.\cite{cotta}. The interesting point
here is that the bipolariton linewidth narrowing never reach the
polariton linewidth, even in the previous measurement, where the
polariton laser emission goes down for larger detuning.

The population inversion is a well known behavior in usual lasers.
One way to characterize it is to observe the kink from spontaneous
to laser emission in a $logxlog$ input/output power measurement
data. The coupling efficiency $\beta$ is related with this kink and
quantifies the efficiency of the laser. In SMC the polariton laser
can operate with almost no population inversion \cite{inversion},
and ß can reach high values as shown by Yamamoto et al.
\cite{inversion2}. Our results for the LB emission showed a larger
kink in the I/O data curve and also in the linewidth data, compared
to our previous behavior observed for the polariton laser
\cite{cotta}. These observations show qualitatively one inversion
population behavior for this bipolariton laser. In the I/O curve
should be possible to quantify $\beta$ by the inverse amplitude of
the kink, however our data intensity resolution were limited by the
spectrum resolution (Fig. \ref{mapa}(a)) and it is possible only to
preview one $\beta$ in the order of previous results for microcavity
laser \cite{cotta}.

The effect of creating the bipolariton emission at high pump powers
is directly associated to the condition of the formation of
biexcitons. This particle is created when there is a high density of
excitons \cite{excitonhigh} in the QW, them by Coulomb interaction
they interact with each other producing excitonic molecules. We
interpreted our results as the following: at high pump power and
resonance condition, the LP scattering is effective, but when the
cavity is detuned, the LP scattering channel decreases the
efficiency, increasing the biexciton population in the QW inducing
thus the bipolariton emission.

We observed in our cavity that the LB emission was blue shifted at
in the PL spectra in Fig. \ref{mapa}, while it was red shifted in
Fig. \ref{potencia}. This could be explained examining the energy
dispersion relation both for polariton and bipolariton in
microcavities. The first one has been extensively studied in the
past few years but in the case of bipolaritons much work has to be
done to fully understand their energy dispersion relation. In a
quantum mechanical approach the energy dispersion relation is
calculated for polaritons in microcavities, where the exciton energy
can be simplified as a constant in range of small wave vectors (our
$k_{\|}$ vector is still a good approximation). The exciton photon
coupling obeying a parabolic energy dispersion form, leads to the
formation of the LP and UP branches. Changing the cavity detuning,
the form of energy dispersion relation of the LP and UP can be
slightly modified \cite{analise}, so depending on the $\Delta$ shift
we are able to change the energy of the transition of the LP and UP
to the valence band, and the resonance condition. In a simplified
model, the bipolariton energy dispersion form is quite similar to
the polariton one. Considering that the biexciton energy is
$E_{xx}(\mathbf{k})=2E_{x}(\mathbf{k})-E^{b}_{xx}$, where
$E_{x}(\mathbf{k})$ is the exciton and $E^{b}_{xx}$ is the biexciton
binding energy. In this equation the biexciton energy is almost
twice the exciton energy, but in fact the recombination process and
photon frequency emitted ($\nu_{xx}$) of the biexciton can be
understand as follows:
\begin{equation}
(biexciton) \rightarrow (exciton) + h\nu_{xx}
\end{equation}
By energy conservation law, we have:
\begin{equation}
h\nu_{xx} = E_{x}(\mathbf{k})- E^{b}_{xx}
\end{equation}
The equations above shows that transition energy of the biexciton is
in fact lower than the exciton recombination energy. For
simplification $E_{xx}$ is also constant in the range of small wave
vectors, then it can couple with two cavity photons and form the
bipolariton upper and lower branches. Also these branches depends on
D, thinking in the UP, LP, UB and LB all together as a function of
the $\Delta$. Now, to interpret our experimental results for the LB
behavior, what would happen is that for some detunings the photon
emitted by the LB can have higher or lower energy with respect to
the LP emission, because the two quasi-particles have different
effective mass, than the form of the energy dispersion relation will
change differently. In Ref. \cite{baars} a calculation of the energy
dispersion for polariton and bipolariton as a function of the
detuning shows that the LP and LB energy can reach the same values.
But this calculation was restricted to some detunings detuning
values and did not show crossing effect on the energy dispersion
relation. Further calculation is calculations are necessary to stand
this interpretation, but it is beyond the present scope of this
paper.

\begin{figure}
\includegraphics[angle = 0, scale = 4.1]{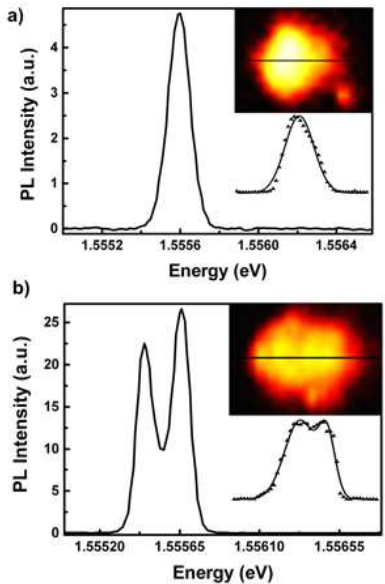}
\caption{(a) PL spectra for the lower polariton emission with the
near field image in the right, and bellow it is the intensity
profile of the marked line where the closed triangles is the
acquired data and the full line is a gaussian fit. (b) The same for
the lower bipolariton emission.} \label{image}
\end{figure}

Finally we measured the near field image with the respective spectra
at 400 mW pump power for the LP only (Fig. \ref{image}(a)) and for
both LP and LB (Fig. \ref{image}(b)) emission by placing a second
camera in front of the collimated beam before the spectrometer.  The
inserted picture in Fig. \ref{image} (b) shows a second pattern for
the bipolariton emission and bellow it the intensity profile of the
image showing clearly that there are two separated beams. Also, both
profiles (LP and LB) are well fitted with a gaussian curve. This
measurement confirms the distinct operation mechanism for the
polariton and the bipolariton laser in the microcavity.

\section{IV.Conclusion}
In conclusion, by detuning the cavity position of the sample for a
fixed energy and high pump power, i.e. detuning the cavity energy in
relation to the exciton, we observed bipolariton emission in a GaAs
planar microcavity. The emission power (I/O) data relation and the
linewidth behavior show one usual laser behavior with population
inversion in the bipolariton emission. The observed bipolariton
emission was red or blue shifted relative to the polariton emission,
depending on the cavity detuning. This switching effect is still a
matter of study to understand the bipolariton scattering process in
a microcavity. The near field images shows that the both emissions
can be not just spectrally separated, but spatially separated as
well.

\section{V.Acknowledgements}
The authors acknowledge financial support from the Instituto de
Nanosci\^encias - MCT, CNPq, Fapemig and CAPES.

\pagebreak

\pagebreak

\end{document}